\documentclass[10pt, twocolumn, pre, aps, superscriptaddress, showpacs]{revtex4-1}
\usepackage{amsmath, graphicx, subfigure, tikz}
\begin{document}

\title{Spinodal and Equilibrium Global Phase Diagram of\\
the d=3 Merged Potts-Cubic-Clock Model:\\
First-Order Equilibrium and Second-Order Spinodal Boundaries with\\
Hidden Topologies from Renormalization-Group Theory}

\author{Umut Açıkel}
\affiliation{Faculty of Engineering and Natural Sciences, Kadir Has University, Cibali, Istanbul 34083, Turkey}
\author{A. Nihat Berker}
    \affiliation{Faculty of Engineering and Natural Sciences, Kadir Has University, Cibali, Istanbul 34083, Turkey}
    \affiliation{Department of Physics, Massachusetts Institute of Technology, Cambridge, Massachusetts 02139, USA}

\begin{abstract}
A model that merges the Potts, cubic, and clock models is studied in spatial dimension d=3 by renormalization-group theory.  Effective vacancies are included in the renormalization-group initial conditions. In the global phase diagram, 5 different ordered phases, namely ferromagnetic, antiferromagnetic, ferrimagnetic, antiferrimagnetic, axial, and a disordered phase are found, separated by first- and second-order phase boundaries. 8 different phase diagram cross-sections occur. When the effective vacancies are suppressed, the global spinodal phase diagram is found: All disordering phase transitions become second order, the disordered phase recedes, and 17 different phase diagram cross-sections occur, spinodality thus much enriching ordering behavior.  In the spinodal phase diagram, the ferrimagnetic and antiferrimagnetic phases have reentrance. The employed renormalization group transformation is exact on the $d=3$ dimensional hierarchical model and Migdal-Kadanoff approximate on the cubic lattice.
\end{abstract}
\maketitle

\section{Merged: Potts, Cubic, Clock Models}

\begin{figure}[ht!]
\centering
\includegraphics[scale=0.40]{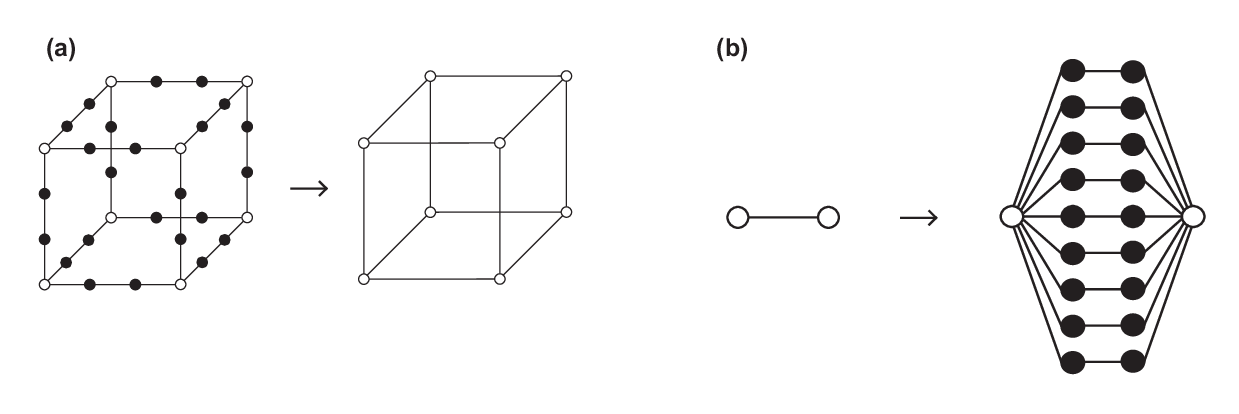}
\caption{Exactly solved $d=3$ hierarchical lattice and Migdal-Kadanoff: (a) The Migdal-Kadanoff approximate renormalization-group transformation on the cubic lattice. Bonds are removed from the cubic lattice to make the renormalization-group transformation doable.  The removed bonds are compensated by adding their effect to the decimated remaining bonds.  (b) A hierarchical model is constructed by self-imbedding a graph into each of its bonds, \textit{ad infinitum}.\cite{BerkerOstlund}  The exact renormalization-group solution proceeds in the reverse direction, by summing over the internal spins shown with the dark circles.  Here is the most used, so called "diamond" hierarchical lattice \cite{BerkerOstlund,Kaufman1,Kaufman2,BerkerMcKay}.  The length-rescaling factor $b$ is the number of bonds in the shortest path between the external spins shown with the open circles, $b=3$ in this case. The volume rescaling factor $b^d$ is the number of bonds replaced by a single bond, $b^d=27$ in this case, so that $d=3$.}
\end{figure}

The mergure of the much used Potts, cubic, clock models is realized by the Hamiltonian
\begin{multline}
- \beta {\cal H} =  \sum_{\left<ij\right>} \, -\beta {\cal H}_{ij}(\vec s_i,\vec t_i;\vec s_j,\vec t_j) = \\
\sum_{\left<ij\right>} \, \{J\, \delta(\vec s_i,\vec s_j)  \, + K \,  [\delta(\vec s_i,\vec s_j)-\delta(\vec s_i,-\vec s_j)]        + C \, \vec s_i \cdot \vec s_j\}
\end{multline}
where $\beta=1/k_{B}T$, at site $i$ the spin $\vec s_i$ can point in $q$ different directions $\theta _i = 2\pi n_i/q$ in the $xy$ plane, with $n_{i}=0,1,...,q-1$ providing the $q$ different possible states, the delta function $\delta(\vec s_i,\vec s_j)=1(0)$ for $\vec s_i = \vec s_j (\vec s_i \neq \vec s_j)$, the last term is a vector product, and the sum is over all interacting pairs of nearest-neighbor spins. We independently vary the Potts interaction strength $J$, the cubic interaction $K$, and the clock interaction strength $C$ of this merged Potts-cubic-clock model, to obtain the multistructured, equilibrium and spinodal, global phase diagram. In this study, $q=6$ is studied.
\begin{figure*}[ht!]
\centering
\includegraphics[scale=0.38]{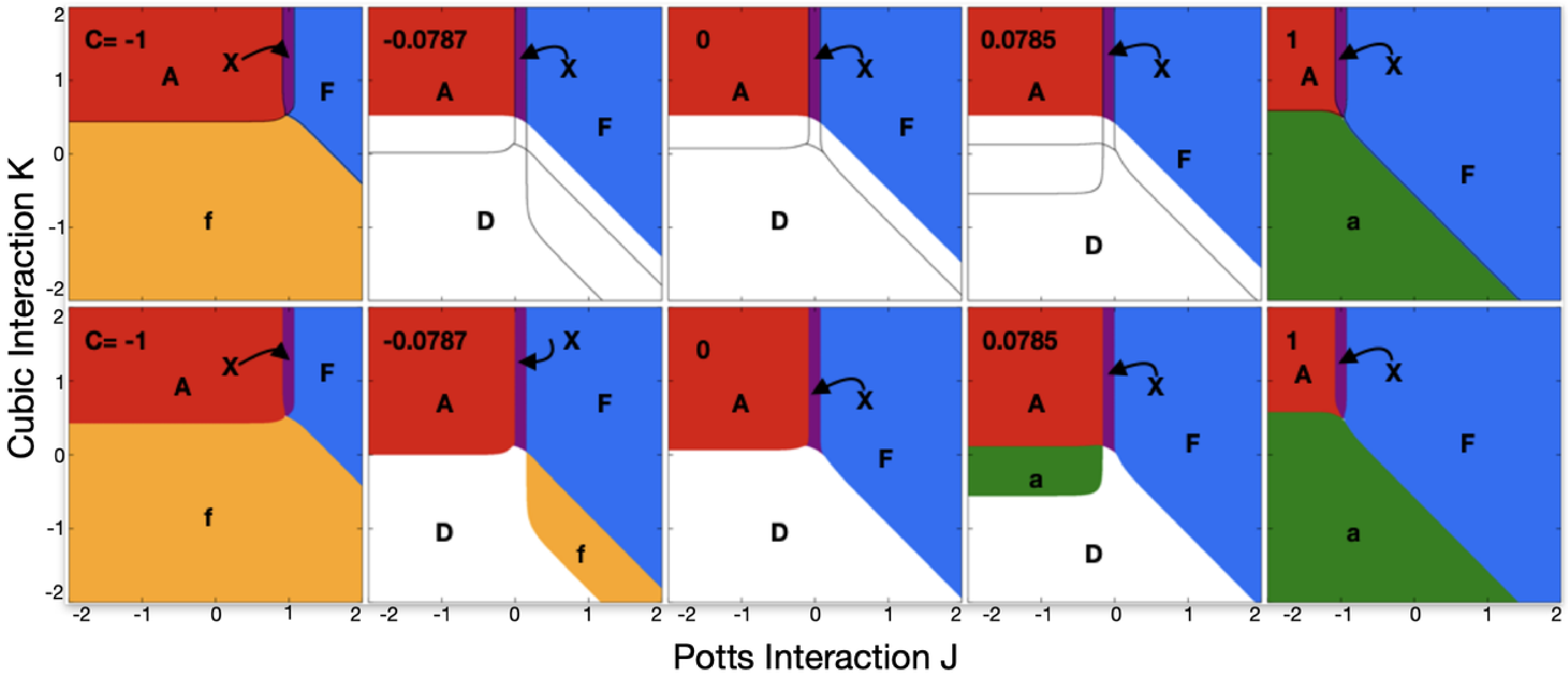}
\caption{Calculated spinodal and equilibrium global phase diagram cross-sections, at various fixed values of the clock interaction $C$, of the Potts-cubic-clock model. These constant $C$ values are given in the upper left of each phase diagram. The ferromagnetic $(F)$, antiferromagnetic $(A)$, ferrimagnetic $(f)$, antiferrimagnetic $(a)$, axial $(X)$ and disordered $(D)$ phases are seen. The top row shows the equilibrium phase diagram. The phase transitions to the disordered phase (D) are first order.  All other phase boundaries are second order.  Three different types of phase diagram topologies, namely the leftmost, middle, rightmost, occur.  The lines show the spinodal boundaries.  The bottom row shows the spinodal phase diagram, obtained by suppressing the effective vacancies. In the bottom spinodal row, the disordered phase has receded and the disordering transitions are second order, while the other phase transitions remain second order.  Five different types of phase diagram topologies appear. }
\end{figure*}

\section{Method: Exact 3d Hierarchical and Physical Migdal Kadanoff, with Condensation of Effective Vacancies}
We use the global renormalization-group theory of exactly solvable $d=3$ hierarchical model and, equivalently, the physically motivated Migdal-Kadanoff approximation, which have been much discussed and much used.  These equivalent procedures consists of decimation and bond moving. The above can be rendered algebraically in the most straightforward way by writing the transfer matrix between two neighboring spins,

\begin{widetext}
\begin{gather}
\textbf{T}_{ij} \equiv e^{-\beta {\cal H}_{ij}} =
\left(
\begin{array}{cccccc}
e^{J+K+C} & e^{0.5C} & e^{-0.5C} & e^{-K-C} & e^{-0.5C} & e^{0.5C} \\
e^{0.5C} & e^{J+K+C} & e^{0.5C} & e^{-0.5C} & e^{-K-C}    & e^{-0.5C}      \\
e^{-0.5C} & e^{0.5C} & e^{J+K+C} & e^{0.5C} & e^{-0.5C}   & e^{-K-C}     \\
e^{-K-C} & e^{-0.5C} & e^{0.5C} & e^{J+K+C} & e^{0.5C}    & e^{-0.5C}     \\
e^{-0.5C} & e^{-K-C} & e^{-0.5C} & e^{0.5C} & e^{J+K+C}   & e^{0.5C}     \\
e^{0.5C} & e^{-0.5C} & e^{-K-C} & e^{-0.5C} & e^{0.5C} & e^{J+K+C}  \end{array} \right),
\end{gather}
\end{widetext}

\begin{figure*}[ht!]
\centering
\includegraphics[scale=0.3]{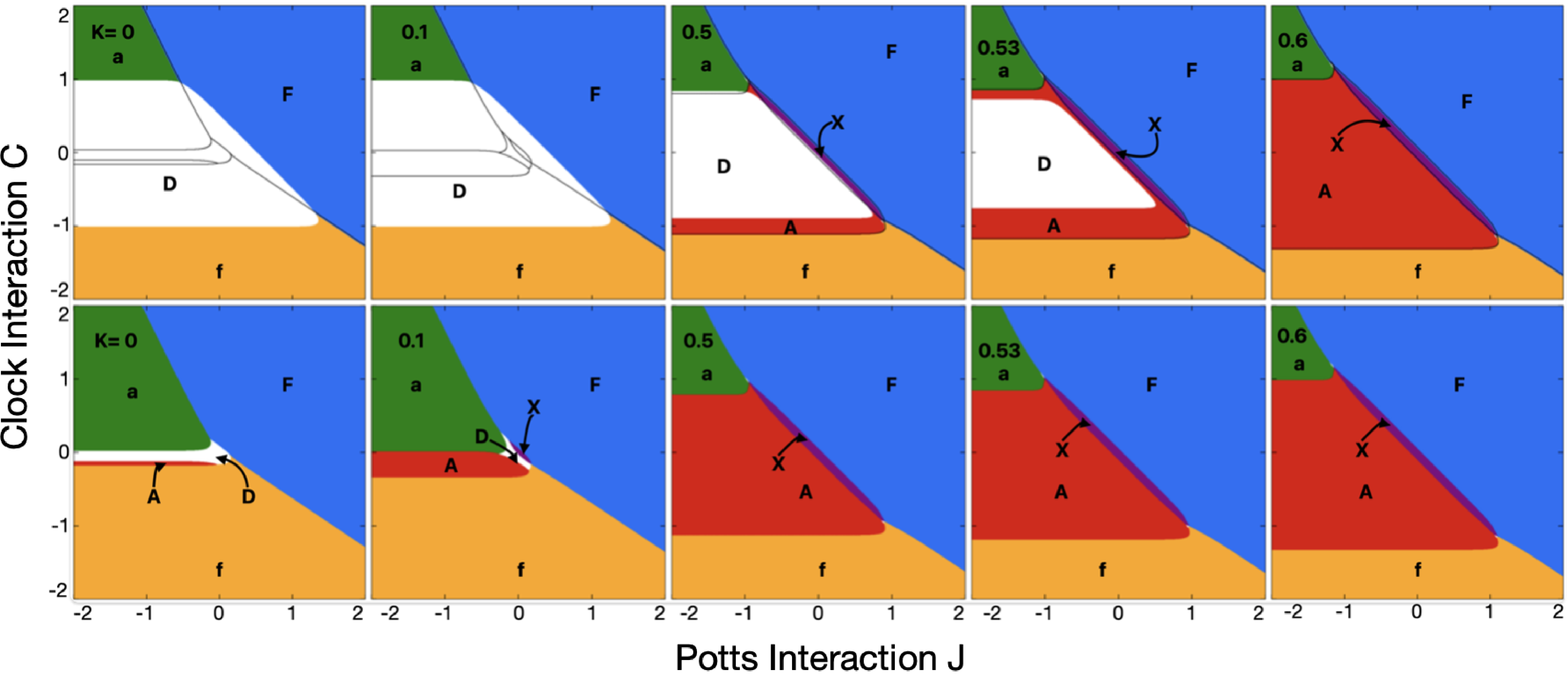}
\caption{Calculated spinodal and equilibrium global phase diagram cross-sections, at various fixed values of the cubic interaction $K$, of the Potts-cubic-clock model.  These constant $K$ values are given in the upper left of each phase diagram. The ferromagnetic $(F)$, antiferromagnetic $(A)$, ferrimagnetic $(f)$, antiferrimagnetic $(a)$, axial $(X)$ and disordered $(D)$ phases are shown. The top row shows the equilibrium phase diagram.   The phase transitions to the disordered phase (D) are first order.  All other phase boundaries are second order.  Three different types of phase diagram topologies, namely the leftmost, middle, rightmost, occur.  The lines show the spinodal boundaries.  The bottom row shows the spinodal phase diagram, obtained by suppressing the effective vacancies. In the bottom spinodal row, the disordered phase has receded and the disordering transitions are second order, while the other phase transitions remain second order.  Three different types of phase diagram topologies occur. }
\end{figure*}

An important aspect of an occurring phase transition is the order of the phase transition. The $q$-state Potts models have a second-order phase transition for $q\leq q_c$ and a first-order phase transition for $q>q_c$.\cite{Baxter,duality1,MonteCarlo} In renormalization-group theory, this has been understood and reproduced as a condensation of effective vacancies formed by regions of local disorder.\cite{spinS7,AndelmanPotts2}
The above has been included \cite{Devre,Kecoglu} as a local disorder state into the two-spin transfer matrix of Eq.(2).  Inside an ordered region of a given spin value, a disordered site does not significantly contribute to the energy in Eq.(1), but has an entropic contribution from a multiplicity of $q-1$.  The substraction comes from the fact that the disordered site cannot be in the spin state of its surrounding ordered region.  This is equivalent to the exponential of an on-site energy and, with no approximation, is shared on the transfer matrices of the $2d$ incoming bonds.  The transfer matrix becomes
\begin{widetext}
\begin{gather}
\textbf{T}_{ij} \equiv e^{-\beta {\cal H}_{ij}} =
\left(
\begin{array}{ccccccc}
e^{J+K+C} & e^{0.5C} & e^{-0.5C} & e^{-K-C} & e^{-0.5C} & e^{0.5C} &(q-1)^{1/2d} \\
e^{0.5C} & e^{J+K+C} & e^{0.5C} & e^{-0.5C} & e^{-K-C}    & e^{-0.5C}  &(q-1)^{1/2d}    \\
e^{-0.5C} & e^{0.5C} & e^{J+K+C} & e^{0.5C} & e^{-0.5C}   & e^{-K-C}   &(q-1)^{1/2d}  \\
e^{-K-C} & e^{-0.5C} & e^{0.5C} & e^{J+K+C} & e^{0.5C}    & e^{-0.5C}  &(q-1)^{1/2d}   \\
e^{-0.5C} & e^{-K-C} & e^{-0.5C} & e^{0.5C} & e^{J+K+C}   & e^{0.5C}  &(q-1)^{1/2d}   \\
e^{0.5C} & e^{-0.5C} & e^{-K-C} & e^{-0.5C} & e^{0.5C} & e^{J+K+C} &(q-1)^{1/2d}\\
(q-1)^{1/2d} &(q-1)^{1/2d} &(q-1)^{1/2d} &(q-1)^{1/2d}   &(q-1)^{1/2d} &(q-1)^{1/2d} &(q-1)^{1/d}  \end{array} \right).
\end{gather}
\end{widetext}

\begin{figure*}[ht!]
\centering
\includegraphics[scale=0.35]{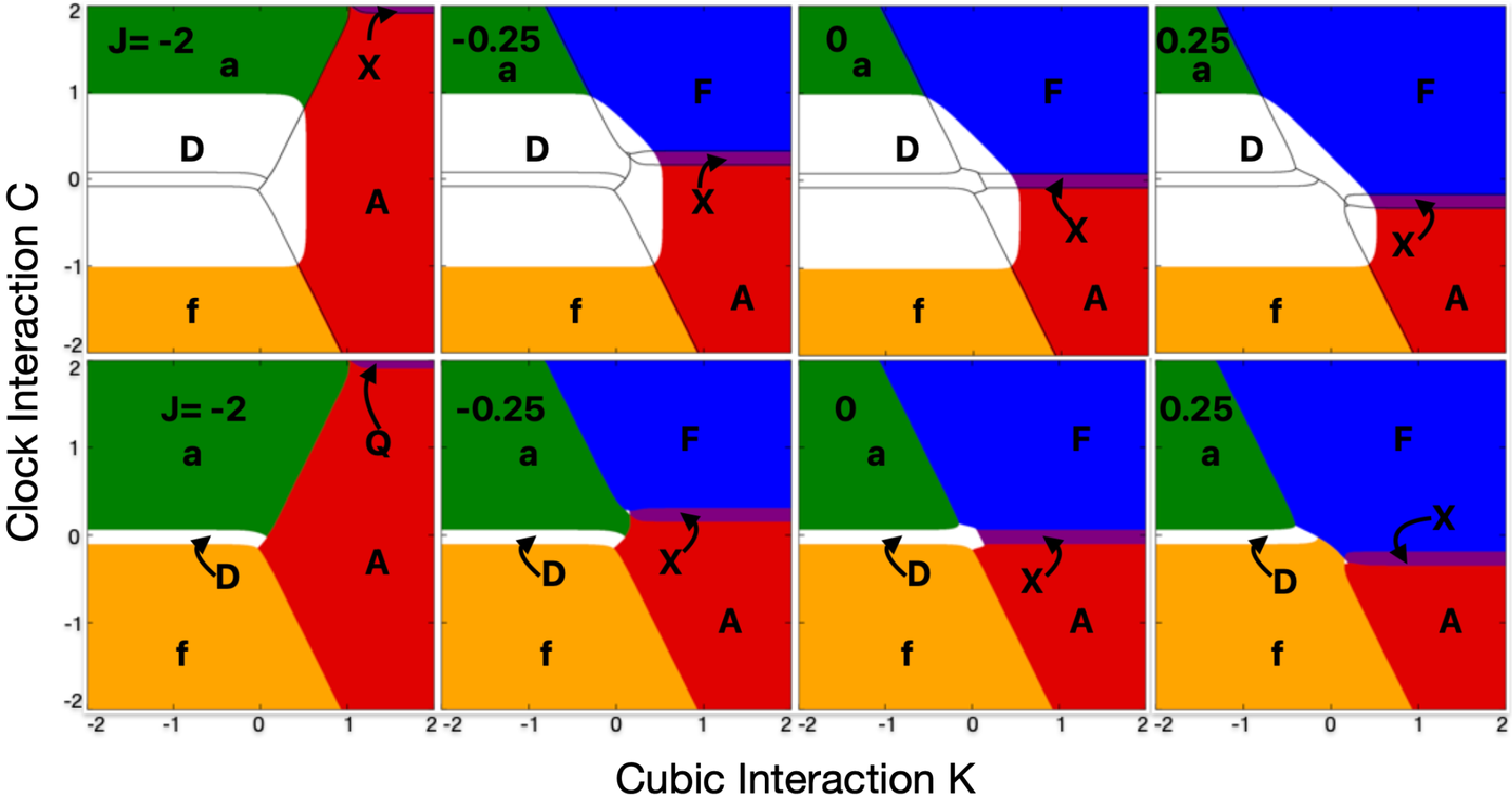}
\caption{Calculated spinodal and equilibrium global phase diagram cross-sections, at various fixed values of the Potts interaction $J$, of the Potts-cubic-clock model.  These constant $J$ values are given in the upper left of each phase diagram. The ferromagnetic $(F)$, antiferromagnetic $(A)$, ferrimagnetic $(f)$, antiferrimagnetic $(a)$, axial $(X)$ and disordered $(D)$ phases are shown. The top row shows the equilibrium phase diagram.   The phase transitions to the disordered phase (D) are first order.  All other phase boundaries are second order.   Two different types of phase diagram topologies, namely the leftmost and rightmost, occur.  The lines show the spinodal boundaries.  The bottom row shows the spinodal phase diagram, obtained by suppressing the effective vacancies. In the bottom spinodal row, the disordered phase has receded and the disordering transitions are second order, while the other phase transitions remain second order.  We note the rich variety of the spinodal phase diagram topologies:  Five different types of phase diagram topologies occur. }
\end{figure*}

\begin{table*}

\begin{tabular}{c}
\multicolumn{1}{c}{Renormalization-Group Sinks of the Thermodynamic Phases} \\
\end{tabular}

\begin{tabular}{c c c c c c c c c c c c}

\hline

\vline & $\begin{pmatrix} 1 & 0 & 0 & 0 & 0 & 0 & 0 \\ 0 & 1 & 0 & 0 & 0 & 0 & 0 \\ 0 & 0 & 1 & 0 & 0 & 0 & 0\\ 0 & 0 & 0 & 1 & 0 & 0 & 0 \\ 0 & 0 & 0 & 0 & 1 & 0 & 0 \\ 0 & 0 & 0 & 0 & 0 & 1 & 0  \\ 0 & 0 & 0 & 0 & 0 & 0 & 0    \end{pmatrix}$ &\vline  & $\begin{pmatrix} 0 & 0 & 0 & 1 & 0 & 0 & 0 \\ 0 & 0 & 0 & 0 & 1 & 0 & 0 \\ 0 & 0 & 0 & 0 & 0 & 1 & 0\\ 1 & 0 & 0 & 0 & 0 & 0 & 0 \\ 0 & 1 & 0 & 0 & 0 & 0 & 0 \\ 0 & 0 & 1 & 0 & 0 & 0 & 0  \\ 0 & 0 & 0 & 0 & 0 & 0 & 0    \end{pmatrix}$ &\vline & $\begin{pmatrix} 1 & 0 & 1 & 0 & 1 & 0 & 0 \\ 0 & 1 & 0 & 1 & 0 & 1 & 0 \\ 1 & 0 & 1 & 0 & 1 & 0 & 0\\ 0 & 1 & 0 & 1 & 0 & 1 & 0 \\ 1 & 0 & 1 & 0 & 1 & 0 & 0 \\ 0 & 1 & 0 & 1 & 0 & 1 & 0  \\ 1 & 0 &1 & 0 & 1 & 0 & 0    \end{pmatrix}$ &\vline &$\begin{pmatrix} 0 & 1 & 0 & 1 & 0 & 1 & 0 \\ 1 & 0 & 1 & 0 & 1 & 0 & 0 \\ 0 & 1 & 0 & 1 & 0 & 1 & 0\\ 1 & 0 & 1 & 0 & 1 & 0 & 0 \\ 0 & 1 & 0 & 1 & 0 & 1 & 0 \\ 1 & 0 & 1 & 0 & 1 & 0 & 0  \\ 0 & 0 & 0 & 0 & 0 & 0 & 0    \end{pmatrix}$&\vline

\\
\hline
\vline & Ferromagnetic (F) &\vline & Antiferromagnetic (A) &\vline & Ferrimagnetic (f) &\vline & Antiferrimagnetic (a)  \vline \\
\hline

\vline & $\begin{pmatrix} 1 & 0 & 0 & 1 & 0 & 0 & 0 \\ 0 & 1 & 0 & 0 & 1 & 0 & 0 \\ 0 & 0 & 1 & 0 & 0 & 1 & 0\\ 1 & 0 & 0 & 1 & 0 & 0 & 0 \\ 0 & 1 & 0 & 0 & 1 & 0 & 0 \\ 0 & 0 & 1 & 0 & 0 & 1 & 0  \\ 0 & 0 & 0 & 0 & 0 & 0 & 0    \end{pmatrix}$ &\vline  &$\begin{pmatrix} 1 & 1 & 1 & 1 & 1 & 1 & 0 \\ 1 & 1 & 1 & 1 & 1 & 1 & 0 \\ 1 & 1 & 1 & 1 & 1 & 1 & 0\\ 1 & 1 & 1 & 1 & 1 & 1 & 0 \\ 1 & 1 & 1 & 1 & 1 & 1 & 0 \\ 1 & 1 & 1 & 1 & 1 & 1 & 0  \\ 0 & 0 & 0 & 0 & 0 & 0 & 0    \end{pmatrix}$ &\vline & $\begin{pmatrix} 0 & 0 & 0 & 0 & 0 & 0 & 0 \\ 0 & 0 & 0 & 0 & 0 & 0 & 0 \\ 0 & 0 & 0 & 0 & 0 & 0 & 0\\ 0 & 0 & 0 & 0 & 0 & 0 & 0 \\ 0 & 0 & 0 & 0 & 0 & 0 & 0 \\ 0 & 0 & 0 & 0 & 0 & 0 & 0  \\ 0 & 0 & 0 & 0 & 0 & 0 & 1    \end{pmatrix}$ &\vline & & \\
\hline
\vline &Axial (X) &\vline & Disordered (D) &\vline & Disordered (D) &\vline & & \\

\hline

\end{tabular}

\caption{Under repeated renormalization-group transformations, the phase diagram points of the phases of the merged Potts-cubic-clock model flow to the sinks shown on this Table giving the exponentiated nearest-neighbor Hamiltonians, namely the transfer matrices.  The designations, $F,A,f,a,X,D$ in Figs. 2-5 are also shown.}
\end{table*}

\begin{table*}

\begin{tabular}{c}
\multicolumn{1}{c}{Examples Renormalization-Group Fixed Points of Phase Transitions between Thermodynamic Phases} \\
\end{tabular}

\begin{tabular}{c c c c c c c c c c c c}

\hline

\vline & $\begin{pmatrix} 1 & 0 & 0 & 0 & 0 & 0 & 0 \\ 0 & 1 & 0 & 0 & 0 & 0 & 0 \\ 0 & 0 & 1 & 0 & 0 & 0 & 0\\ 0 & 0 & 0 & 1 & 0 & 0 & 0 \\ 0 & 0 & 0 & 0 & 1 & 0 & 0 \\ 0 & 0 & 0 & 0 & 0 & 1 & 0  \\ 0 & 0 & 0 & 0 & 0 & 0 & 1    \end{pmatrix}$ &\vline  & $\begin{pmatrix} 1 & A & B & C & B & A &\\ A & 1 & A & B & C & B &\\ B & A & 1 & A & B & C &\\ C & B & A & 1 & A & B &\\ B & C & B & A & 1 & A &\\ A & B & C & B & A & 1 &\end{pmatrix}$ &\vline & $\begin{pmatrix} 1 & 0 & 0 & 0.924 & 0 & 0 & \\ 0 & 1 & 0 & 0 & 0.924 & 0 & \\ 0 & 0 & 1 & 0 & 0 & 0.924 & \\ 0.924 & 0 & 0 & 1 & 0 & 0 &  \\ 0 & 0.924 & 0 & 0 & 1 & 0 &  \\ 0 & 0 &0.924 & 0 & 0 & 1 &    \end{pmatrix}$ &\vline

\\

\hline

\vline &Equilibrium First-Order &\vline & Spinodal Second-Order &\vline & Equilibrium Second-Order Phase Transiton &\vline \\

\hline

\vline &Ferromagnetic (F) - Disorder (D) &\vline & Ferromagnetic (F) - Disorder (D) &\vline & Ferromagnetic (F) - Axial (X) &\vline \\

\hline

\end{tabular}

\caption{These fixed points attract, under renormalization-group flows, their respective phase boundaries. Under renormalization-group flows, these fixed points have one unstable (namely outflowing) direction, into the two phases they bound, respectively in the opposite flow directions. The calculated values are $A=0.96156,B=0.88572,C=0.84832$.}
\end{table*}

\begin{figure*}[ht!]
\centering
\includegraphics[scale=0.5]{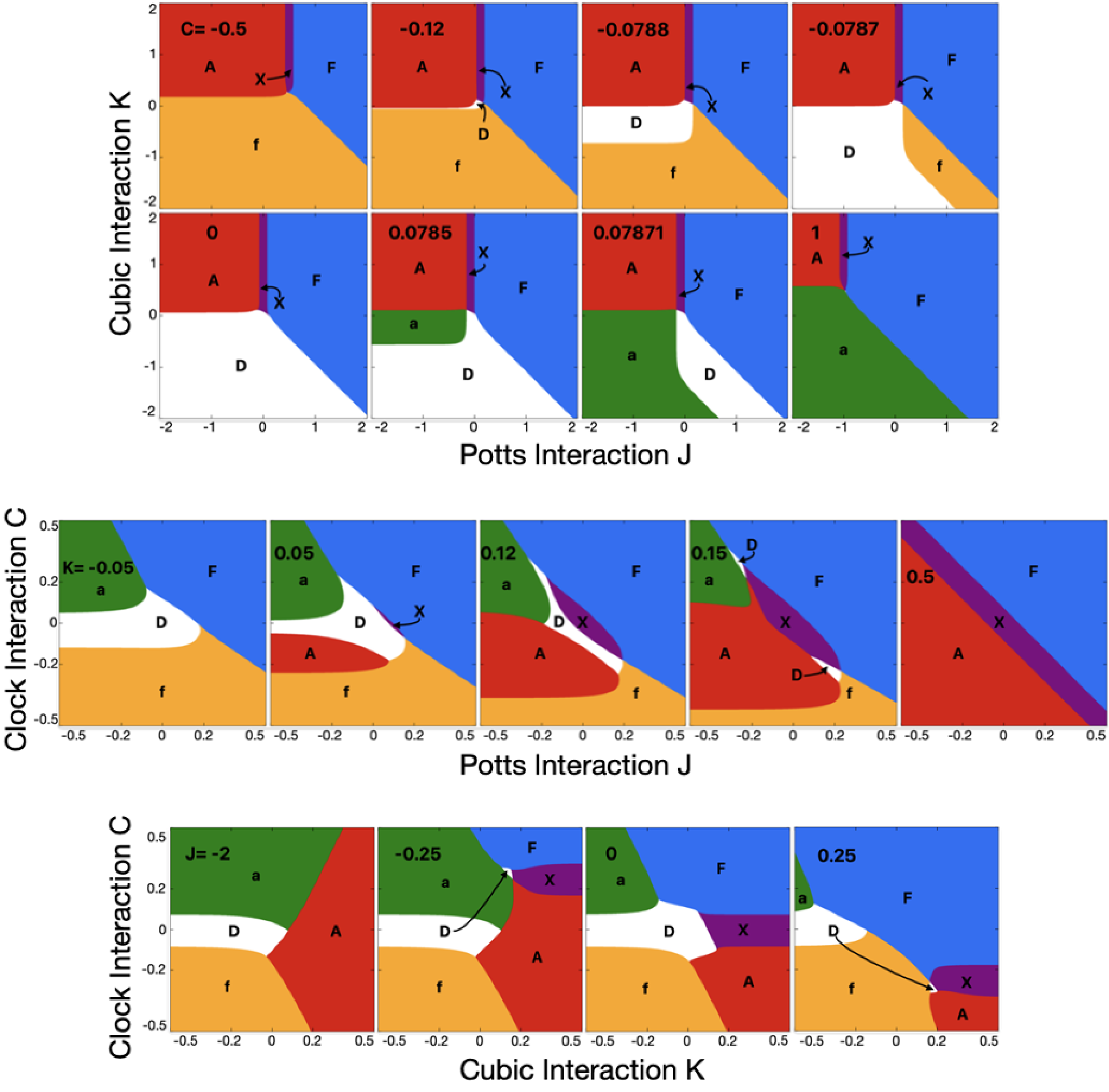}
\caption{Topologically distinct cross-sections of the spinodal global phase diagram The all phase transitions are second order. 17 different phase diagram topologies are seen.}
\end{figure*}

The renormalization-group transformation, explained in Fig. 1, is done with length rescaling factor $b=3$ in order to conserve the ferromagnetic-antiferromagnetic symmetry of the method.  This method \cite{Migdal,Kadanoff} involves decimating three bonds in series into a single bond, followed by bond-moving by superimposing $b^{d-1}=9$ bonds. After each algebraic operation, the transker matrix is divided by its largest element, which amounts to applying a subtractive term to every energy, not affecting the physics but preventing computing overflows.  This approach is an approximate solution on the $d=3$ cubic lattice and, simultaneously, an exact solution on the $d=3$ hierarchical lattice \cite{BerkerOstlund,Kaufman1,Kaufman2,BerkerMcKay}.  The simultaneous exact solution makes the approximate solution a physically realizable, therefore robust approximation, as also used in turbulence \cite{Kraichnan}, polymer \cite{Flory}, gel \cite{Kaufman}, electronic system \cite{Lloyd} calculations. A physically realizable approximation means that the approximation is exact on an alternate physically realizable system and therefore will obey all the general laws of thermoynamics.

Thus, additional to the exactness on the $3d$ hierarchical lattice, a simple but effective, physically inspired (see Fig. 1(a)) method in studying phase transitions has been this renormalization-group method under the Migdal-Kadanoff approximation \cite{Migdal,Kadanoff}.  Thus, using this method on widely different systems, the lower-critical dimension $d_c$ below which no ordering occurs has been correctly determined as $d_c=1$ for the Ising model \cite{Migdal,Kadanoff}, $d_c=2$ for the XY \cite{Jose,BerkerNelson} and Heisenberg \cite{Tunca} models, and the presence of an algebraically ordered phase has been seen for the $d=2$ XY model \cite{Jose,BerkerNelson,Sariyer}.  In systems with frozen microscopic disorder (quenched randomness), using the simple Migdal-Kadanoff renormalization-group approximation on quenched random distributions, $d_c=2$ has been determined for the random-field Ising \cite{Machta,Falicov} and XY models \cite{Kutay}, and, yielding a non-integer value, $d_c=2.46$ for Ising spin-glass systems \cite{Atalay}. Also under the Migdal-Kadanoff approximation, the chaotic nature of the spin-glass phases \cite{McKayChaos,McKayChaos2,BerkerMcKay} has been obtained and quantitatively analyzed, both for quenched randomly mixed ferromagnetic-antiferromagnetic spin glasses \cite{Ilker1,Ilker2,Ilker3} and right- and left-chiral (helical) spin glasses \cite{Caglar1,Caglar2,Caglar3}. The Migdal-Kadanoff procedure has also been used in the calculation of the phase diagrams of surface systems \cite{BerkerPLG}, accurately matching experiments with no adjustable parameter, and finite-temperature phase diagrams of high-temperature superconductivity models \cite{highTc}.  For recent works on hierarchical lattices, see Refs.\cite{Sponge,CubicSG,Clark,Kotorowicz,ZhangPP,Jiang,Derevyagin2,Chio,Teplyaev,Myshlyavtsev,Derevyagin,Shrock,Monthus}

\section{Spinodal and Equilibrium Global Phase Diagrams}

The phase diagrams are obtained by following the renormalization-group trajectories to their terminus of totally attractive fixed ponts, namely sinks, shown in Table I.\cite{BerkerWor}.  The basin of attraction of each fixed point is a thermodynamic phase.  A sink epitomizes the thermodynamic phase that it attracts.  The seven sinks subtending this global phase diagram are given in Table I.  The disordered phase has two sinks, dense disordered and dilute disordered, but no phase transition is correctly seen between them, since this system does not contain physical vacancies.\cite{BerkerPLG}

The fixed point attracting the phase boundary between two thermodynamic phases determines order of the phase transition and, in the case of a second-order phase transition, the critical exponents.\cite{BerkerWor}  Examples are shown in Table II.  The fixed point of a first-order phase transition is the superposition of the sinks of the phases that are bounded.  From the calculated values in the fixed point of a second-order phase transition, the derivative matrix (\textit{aka}, the recursion matrix) is calculated.  For example, for the phase transition between the ferromagnetic and disordered phases in Table II,
\begin{widetext}
\begin{gather}
\begin{pmatrix} \partial A'/\partial A & \partial A'/\partial B & \partial A'\partial C \\ \partial B'/\partial A & \partial B'/\partial B & \partial B'\partial C \\ \partial C'/\partial A & \partial C'/\partial B & \partial C'\partial C\end{pmatrix} = \begin{pmatrix} -0.12001 &-0.50364, &-0.75268 \\  0.56100  &1.62461  &2.12328 \\ 0.28128 & 0.88026 & 1.19962\end{pmatrix}
\end{gather}
\end{widetext}
The eigenvalues of this matrix are $\lambda_1=2.61438, \lambda_2=0.08618,  \lambda_3= 0.00366  $.  The latter two eigenvalues are, as expected, less than unity (irrelevant), their right eigenvectors corresponding to renormalization-group flow attractive directions.  The leading eigenvalue is greater than unity, relevant, its right eigenvector pointing on either side to the bounded phases.  The eigenvalue exponent,
\begin{equation}
y_1 = \ln(\lambda_1) / \ln(b) = 0.874,
\end{equation}
yields the correlation-length critical exponent,
\begin{equation}
\nu = 1/y_1 = 1.14.
\end{equation}

Figure 2 shows the calculated spinodal and equilibrium global phase diagram cross-sections, at various fixed values of the clock interaction $C$, of the Potts-cubic-clock model. The ferromagnetic $(F)$, antiferromagnetic $(A)$, ferrimagnetic $(f)$, antiferrimagnetic $(a)$, axial $(X)$, and disordered $(D)$ phases are seen.  Since a sink epitomizes the thermodynamic phase that it attracts, it is seen that, in the axial phase, the spins preferentially point in $\pm a$ spin-space directions.  Thus this phase is threefold degenerate and its ground state has an entropy of $\ln 2$ per bond.  The ferrimagnetic phase has spins pointing in three spin-space directions separated by $2\pi /3$, is therefore doubly degenerate and with ground-state entropy of $\ln 3$ per bond.  The antiferrimagnetic phase forms two such sublattices, is also doubly degenerate and with ground-state entropy of $\ln 3$ per bond.  The ferromagnetic and antiferromagnetic phases show reentrance \cite{Cladis,Netz,Garland,Walker,Caflisch,transverse,Ilker1,Mann1,Mann2}, meaning the disappearance and reappearance of a phase, when proceeding in the phase diagram along a straight line emerging from infinite temperature $(J=M=0)$.

The top row ın Fig. 2 shows the equilibrium phase diagram.  The phase transitions to the disordered phase (D) are first order.  All other phase boundaries are second order.  Three different types of phase diagram topologies, namely the leftmost, middle, rightmost, occur.  The superimposed lines in the top row show the spinodal boundaries.  The bottom row shows the spinodal phase diagram, obtained by suppressing the effective vacancies. The disordered phase recedes and the disordering transitions are second order, while the other phase transitions remain second order.  Five different types of phase diagram topologies appear.

Figure 3 shows the calculated spinodal and equilibrium global phase diagram cross-sections, at various fixed values of the cubic interaction $K$, of the Potts-cubic-clock model. The top row shows the equilibrium phase diagram.  The phase transitions to the disordered phase (D) are first order.  All other phase boundaries are second order.   Three different types of phase diagram topologies, namely the leftmost, middle, rightmost, occur.  The superimposed lines in the top row show the spinodal boundaries.  The bottom row shows the spinodal phase diagram, obtained by suppressing the effective vacancies. The disordered phase recedes and the disordering transitions are second order, while the other phase transitions remain second order.  Three different types of phase diagram topologies occur.

Figure 4 shows the calculated spinodal and equilibrium global phase diagram cross-sections, at various fixed values of the Potts interaction $J$, of the Potts-cubic-clock model. The top row shows the equilibrium phase diagram.  The phase transitions to the disordered phase (D) are first order.  All other phase boundaries are second order.   Two different types of phase diagram topologies, namely the leftmost and rightmost, occur.  The superimposed lines in the top row show the spinodal boundaries.  The bottom row shows the spinodal phase diagram, obtained by suppressing the effective vacancies. The disordered phase recedes and the disordering transitions are second order, while the other phase transitions remain second order.  Four different types of phase diagram topologies occur.

It is thus seen that the spinodal global phase diagram has a much richer phase boundary structure than the equilibrium global phase diagram.  The rich phase boundary structure of the spinodal phase diagram gets preempted by the first order transitions under equilibrium.  The totality of the spinodal phase diagrams is seen in Fig. 5, exhibiting the phase reentrances \cite{Cladis,Netz,Garland,Walker,Caflisch,transverse,Ilker1,Mann1,Mann2} of the ferrimagnetic and antiferrimagnetic phases.

\section{Conclusion}

We have solved, by renormalization-group theory, the merged Potts-cubic-clock model, obtaining the spinodal and eqilibrium global phase diagrams. In the equilibrium global phase diagram, 5 different ordered phases and a disordered phase are found, separated by first- and second-order phase boundaries. 8 different phase diagram cross-sections occur. In the spinodal global phase diagram, the disordering phase transitions become second order, the order-to-order phase transitions remain second order, the disordered phase recedes, reentrances of two separate phases appear, 17 different phase diagram cross-sections occur. Suppression of the vacancies has been used to obtain the spinodal phase diagram from renormalization-group theory, yielding the richer phase diagram from under the receding disordered phase.

\begin{acknowledgments} Support by the Academy of Sciences of Turkey (T\"UBA) is gratefully acknowledged.
\end{acknowledgments}

\end{document}